\documentclass[preprint,review,12pt]{elsarticle}

\usepackage{epsfig}

\usepackage{amssymb}

\journal{Int. J. of Thermal Sciences}

\begin{document}

\begin{frontmatter}


\title{Thermal Response of Dielectric Nanoparticle Infused Tissue Phantoms during Microwave Assisted Hyperthermia}


\author{Dhiraj Kumar$^{1}$}
\author{Purbarun Dhar$^{2}$}
\author{Anup Paul$^{1,}$\corref{corl}}
\cortext[corl]{email: catchapu@gmail.com, 
Tel.: +91-0360-2284801, 
Fax: +91-0360-2284972}


\address{$^{1}$Department of Mechanical Engineering,
National Institute of Technology Arunachal Pradesh, Yupia, Arunachal Pradesh-791112, India\\
$^{2}$Department of Mechanical Engineering,
Indian Institute of Technology Kharagpur, Kharagpur-721302, West Bengal, India}

\begin{abstract}
Hyperthermia has been in use for many years; as a potential alternative modality for cancer treatment. In this paper, an experimental investigation of microwave assisted thermal heating (MWATH) of tissue phantom using a domestic microwave oven has been reported. Computer simulations using finite element method based tools was also carried out to support the experimental observations and probe insight on the thermal transport aspects deep within the tissue phantom. A good agreement between predicted and measured temperature were achieved. Furthermore, experiments were conducted to investigate the efficacy of dielectric nanoparticles viz. alumina (Al$_{2}$O$_{3}$) and titanium oxide (TiO$_{2}$) during the MWATH of nanoparticle infused tumor phantoms. A deep seated tumor injected with nanoparticle solution was specifically mimicked in the experiments. Interesting results were obtained in terms of spatiotemporal thermal history of the nanoparticle infused tissue phantoms. An elevation in the temperature distribution was achieved in the vicinity of the targeted zone due to the presence of nanoparticles, and the spatial distribution of temperature was grossly morphed. We conclusively show, using experiments and simulations that unlike other nanoparticle mediated hyperthermia techniques, direct injection of the nanoparticles within the tumor leads to enhanced heat generation in the neighboring healthy tissues.  The inhomogeneity of the hyperthermia event is evident from the local occurrence of hot spots and cold spots respectively. The present findings may have far reaching implications as a framework in predicting temperature distributions during MWA.
\end{abstract}




\end{frontmatter}


\section{Introduction}
During the last few decades, development of medical devices and techniques for numerous diagnostics and therapeutics has been aided by the advancement of electromagnetics and modern electronics. Many ailments, such as spinal nerve disorders, gum infections, cancer, arthritis, ulcers, and chronic pain can be treated by localized application of electromagnetic field. Today, cancer has become a global threat and currently, one in six disease mediated deaths in the world is due to cancer \citep{who2018}. Cancer may affect people of all ages, even fetuses, but the risk for most varieties increases with age. The cancer of the liver is considered to rank fourth among cancer related deaths and accounted for 782,000 deaths in 2018 \citep{who2018}. Surgery is considered as the primary treatment, but is mostly possible in the early stages of the disease, and that too for patients with limited cirrhosis \citep{Bruix2016Apr.2016Jan12}. During the last decade, Radiofrequency (RF) and Microwave ablation (MWA) have surfaced as important non-surgical and minimally-invasive hyperthermia treatment for liver cancer \citep{Verslype2012Oct}. 

Hyperthermia is a medical treatment where the whole body or an infected region is exposed to high temperatures to either damage or even exterminate cancerous cells. The modality can also be used to make cancer cells more sensitive to radiation and certain drugs (along with chemotherapy). Such thermal approaches are minimally or non-invasive, relatively simple to perform, and have the potential of treating embedded tumors in vital regions where surgery is not feasible or risky. However, in order to reach underlying tumors or thermally ablate large tumors, the activating energy source must sufficiently penetrate through healthy tissues, while minimizing damage to the later. Unfortunately, simple heating techniques have trouble discriminating between tumors and surrounding healthy tissues, and often lead to heat damage of the tissue between the source and the target site. The MWA, which was first introduced in 1994 \citep{Seki1994Aug1}, has gained importance for treatment of liver cancer due to its several significant clinical advancements. These advancements include improvements in microwave antenna design, spatial and concurrent distribution of power to multiple antennas, \citep{Laeseke2009, Harari2016Jan} and the development of multi slot antennas with internal cooling technology \citep{knavel2012, Poulou2015May18}. These advancements help the MWA modality to achieve higher temperatures and faster heating compared to other hyperthermia modalities (laser, ultrasound etc.). 

The major advantages of MWA over other external heating sources are shorter ablation times, larger necrosis volumes, and negligible heat sink effect due to major blood vessels \citep{Poulou2015May18}. During MWA, an electromagnetic wave of either 915 or 2450 MHz is applied to the target site using microwave applicator, heating it to $>$150 $^{o}$C via dielectric polarization \citep{Kim2018Jul}. The technique is most effective for tissues with high water content \citep{Poulou2015May18}. Theoretical modeling and/or computer simulations of hyperthermia therapy play important roles in successful development and implementation by providing a-priori information during the clinical therapy. The computer simulations serve as powerful tools by predicting precise necrotic temperature, ablation volume, time of heating, thermal history and distribution, and tissue thermal damage. From literature it is evident that several simulation studies \citep{Prakash2010Feb4, Chiang2013Jun, Lopresto2017Feb, Cavagnaro2015Apr21, BARAUSKAS2008694, Lu2009Mar, Liu2017Mar21, XU2019539, doi:10.1002/mmce.21581, Yang2007Jan, Deshazer2017Feb} exist on the modeling of thermotherapy and the efficacy of MWA targeted to different tumor bearing organs.

Notably, Cavagnaro et al. \citep{Cavagnaro2015Apr21} have proposed different numerical models in terms of temperature dependent thermal and dielectric properties on the temperature distribution and thermal ablation volume. In a pioneering work by Lu et al. \citep{Lu2009Mar}, the temperature distribution has been predicted using finite element based simulations using water cooled MWA antenna. Xu et al. \citep{XU2019539} numerically predicted the temperature distribution within liver tissue using periodically distributed tri-slot microwave antenna. Some studies reported \citep{Seki1994Aug1, Laeseke2009, Harari2016Jan, knavel2012, Poulou2015May18, Kim2018Jul, Dong1998Aug, Brace2009MayJun} experimental findings on the thermal ablation of tumors during MWA, mostly for treating percutaneous local treatment of small volume liver cancer. To achieve uniform temperature distribution and precise tumor ablation, multiple antenna configurations with simultaneous power delivery is considered to be less invasive \citep{Laeseke2009, Harari2016Jan, Brace2009MayJun} than single antenna configuration with sequential power delivery. During MWA using multiple antennas, a single generator can supply microwave power to multiple antennas arranged in array. This leads to ablation of large volume of tissue along with coagulation of large blood vessels \citep{Brace2007Jul, Wright2003Apr}. Ping Liang et al. \citep{930907} studied double-applicator configurations, Brace et al. \citep{Brace2007Jul} and Ryan et al. \citep{Ryan2010} proposed antennas arranged in triangular arrays, and Phasukkit et al. \citep{Phasukkit2009Nov} proposed different geometrical configurations, all for targeted MWA of hepatic cancers. Specifically, Ryan et al. \citep{Ryan2010} have shown that the treatment efficiency can be improved by operating antennas synchronously. However, in most of the studies reported, the heating time has been considered to be 2-10 minutes, which results in overheating of the surrounding tissue due to heat diffusion. 

The studies conducted on numerical and experimental investigation of MWA of tissues focus on using different microwave antennas. However, the antennas are very costly and hence make the research endeavor expensive. Consequently, a detailed understanding of thermal history and distribution in tissues during MWA is yet to be achieved. Also, to ablate large tissue volumes, multiple antennas placed in an array is required, which in turn results in overheating of the surrounding healthy tissue. Experimental understanding of the same is also absent due to the cost and non-clinical availability of such arrays. Further, the aspects of thermal damage to neighboring healthy tissues during MWA require strong attention to minimize post-treatment loss of health and alleviate patient comfort levels. In recent years, the microwave oven (a common kitchen appliance) has gained momentum in research involving heating of biomass and biomaterials \citep{HALIM20191191, datta1993, doi:10.1111/j.1365-2621.1993.tb06198.x, NI19991501, SALEMA201512, CLARK2000153}, of nanocolloids \citep{JACOB201245}, and of other types of materials \citep{MIMOSO2017175, CHERBANSKI2013214, KUMAR2016304, KAPRANOV20102319} Since the domestic oven operates at 2450 MHz (the most common frequency for MWA), it provides an inexpensive yet very realistic method to study MWA. 

In this paper, detailed experimental and computational simulations have been presented to understand the thermal history and distribution within equivalent tissue phantoms during MWA. The study involves the use of a domestic microwave oven for the experimentations, including real time thermal monitoring. The simulations and experiments are conglomerated to provide a realistic model and validation of the thermal states of tissue phantoms during microwave based hyperthermia. The paper then tackles the concern of thermal damage to neighboring healthy tissues by the use of nanotechnology. The region within the phantom which mimics the tumor is infused with dielectric nanoparticles. The thermo-generation aspect of the nanoparticles due to dielectric resonance is appealed to and enhanced heating of the tumor site, reduced hyperthermia time, and reduced damage to neighboring healthy tissue is achieved. Here also, the experiments are supplemented with simulations and very good match is noted between the two approaches. The present study may have strong implications towards development of patient-to-patient based guidelines and MWA hyperthermia protocols by clinicians prior to real clinical diagnostics and therapy.

\section{Experimental methods}
\subsection{Preparation of tissue phantom}
The tissue phantom used was synthesized from agar gel. The gel prepared by dispersing agar powder (Sisco Research Labs, India) in warm ($\backsim$40 $^{o}$C) deionized double distilled water. For optimum mechanical and thermal property to mimic human tissues, the powder concentration used was kept 2.6\% weight by volume \citep{PAUL201477}. Initially, 600 ml of water was heated to 40 $^{o}$C for 3 min (at 200 W) using an induction heater and the temperature was monitored with a thermometer. Next, the agar powder was dissolved in the warm water. The solution was then heated for 10 min and continuously stirred to prevent burning of agar gel, till the murky solution became clear and started boiling. The solution was poured into a 400 ml glass beaker (70 mm inside diameter, Borosil) up to a height of 70 mm, which was pre marked as shown in Fig. 1(a). The glass beaker acts as the test section, with details illustrated in Fig. 1. The bottom surface of the beaker was covered with aluminum foil to avoid loss of moisture from the gel during cooling. Finally, the gel was allowed to cool and solidify at room temperature for 10 hours.

\subsection{Preparation of nanoparticle infused tissue phantom}
Nanoparticles with agar gel mixture were prepared with the particle concentration of 1\% (w/v), in order to mimic MWA of tumors injected with dielectric nanoparticles. The nanoparticle-gel mixture was placed at the center of the phantom sample, and had dimensions of 8 mm $\times$ 10 mm (diameter $\times$ height). Two different nanoparticles (both procured from Sisco Research Labs, India) were used viz., Aluminum Oxide (Al$_{2}$O$_{3}$) with average size of 15 nm, and Titanium Oxide (TiO$_{2}$) with average size of 7 nm. The nanoparticles were chosen as both the materials are good dielectrics, with dielectric constants of Al$_{2}$O$_{3}$ and TiO$_{2}$ as $\backsim$10-12 and 100-120, respectively. For the synthesis of nanoparticle mixed tissue phantom, agar powder (2.6\% w/v) and nanoparticle (1\% w/v) was added to warm ($\backsim$40 $^{o}$C) water. The uniform mixing of nanoparticles was ensured by 30 min stirring using magnetic stirrer. The mixture was then heated to a temperature of 80 $^{o}$C for 10 min, with constant stirring. The mixture is then poured into the sample cavity (8 mm $\times$ 10 mm) in the agar tissue phantom as shown in Fig. 1(b). The spatial coordinate of the nanoparticles infused tissue phantom is shown in Fig. 2(c). Finally, the second layer of agar gel were poured into the test section and allowed to cool to room temperature for 10 hours, thereby sealing the nanoparticle infused tumor phantom within the bulk phantom. \\
\begin{center}
$<$Figure 1$>$
\\
$<$Figure 2$>$
\end{center}

\subsection{Procedure for microwave heating experiments}
Heating of the tissue phantoms was carried out in a microwave oven (2450 MHz frequency, Samsung, India) at power level of 900 W, for continuous heating period of 20 seconds. The glass beaker with the phantom within was the test section. Glass absorbs very less amount of microwaves, thus allowing the phantom to be exposed to majority of the incident radiation. A Teflon sheet was used to fabricate an arrangement for housing thermocouples for temperature measurement at different radial and axial positions. The slotted Teflon lid was placed on top of the phantom surface, equipped with the radially arranged thermocouples to measure the temperature within the phantom at different radial and axial positions, as shown in Fig. 2(c). Teflon is employed as its insulating property, high melting point, and negligible microwave radiation absorbance ensures minimal aberrations during experiments. The schematic of the experimental set up is shown in Fig. 1(d). Precision fine wire J type thermocouples were used to measure temperature and were calibrated using a constant temperature bath (Cole-Parmer, USA), for a temperature range of 20 $^{o}$C to 130 $^{o}$C. The maximum error observed was 0.3 $^{o}$C.

The thermocouples were fixed at four radial positions of 0 mm (center), 10 mm, 20mm and 30 mm, respectively, from the center of the Teflon cover.  The thermocouples were marked at different axial positions for temperature measurement at different depths, and were kept fixed on a wooden fixture (as shown in Fig. 1(c)) with Teflon tape to avoid any displacement during microwave heating. The oven was set to microwave mode, at 900 W, and time of heating was fixed to 20 sec. The temperature data was collected using a data acquisition module (Keysight, USA). The same steps were repeated for each sample, after fixing the thermocouples at different axial and radial positions. For each sample, the experiments were repeated 5 times to determine the uncertainty in the temperature measurement.  The measurement of thermal properties of the tissue phantom was done using a thermal property analyser (TEMPOS, Meter Group, Inc., USA), and was found very identical \citep{Rossmanna2014, YANG2004445} with biological tissues, as shown in Table 1.\\
\begin{center}
$<$Table 1$>$
\end{center}

\section{Computational details}
\subsection{Computational domain}
To predict the thermal history and contour within the tissues, a computational domain is employed to mimic the phantom. It is modeled as a homogenous cylindrical block, neglecting the microvasculature and pores to mimic the actual experimental phantom. The internal cavity of the microwave oven is modelled as a hollow metallic cuboid, connected to a 2.450 GHz microwave source via a rectangular waveguide operating in the TE$_{10}$ mode. On the floor of the oven cavity is a circular glass plate (turn table) with the cylindrical tissue phantom seated on top. The oven cavity has dimensions of 350 mm $\times$ 210 mm $\times$ 350 mm. The computational domain is shown in Fig. 2(a). All the dimensions of the computational model are considered as per the experimental setup and are provided in Table 2. In case of nanoparticle infused tumor, the dimensions can be seen in Fig. 2(c).\\
\begin{center}
$<$Table 2$>$
\end{center}
 
\subsection{Governing equations}
To predict the thermal response of tissue phantoms during microwave heating, the transient heat diffusion equation is solved with initial and boundary conditions. For simplification the following assumptions are resorted to 
\begin{enumerate}[(a)]
\item Microwave radiation is incident on the tissue phantom in a radially symmetrical pattern,
\item The air flow inside the cavity is neglected,
\item Constant dielectric and thermal properties are considered,
\item The tissue phantom is considered to be isotropic and homogenous,
\item Phase change and chemical reactions in the tissue phantom is neglected,
\item Simulation of microwave radiation emission from the waveguide into the cavity is not considered

\end{enumerate}
\subsubsection{Thermal transport model}
The incident microwaves are converted into thermal energy based on the electromagnetic field distribution at a particular location. The absorbed energy is modelled as a source term in the thermal energy equation to calculate the transient temperature profiles. The heat diffusion equation for the tissue phantom is expressed as
\begin{equation}
\rho C_{p}\frac{\partial T}{\partial t}=\bigtriangledown.\left( k_{th} \bigtriangledown T\right) + Q
\end{equation}
where Q is the heat source term in (W/m$^{3}$), $\rho$ is the density (kg/m$^{3}$), C$_{p}$ is the specific heat capacity (J/kg K), k$_{th}$ is the thermal conductivity of the tissue phantom (W/m K). The source term represents the thermo-generation due to the electric field (Q$_{e}$) and magnetic field (Q$_{m}$) components of microwave radiation, and is expressed as \citep{HALIM20191191}
\begin{equation}
Q=Q_{e}+Q_{m}
\end{equation}
For non-magnetic materials such as tissue phantoms,
\begin{equation}
Q=Q_{e}
\end{equation}
\subsubsection{Microwave heating model}
The solution of the coupled electromagnetic waveform of Maxwell's equations provides the estimated peak electric field strength (E) at any point in the computational domain. The electromagnetic and thermal transport equations have been solved numerically. The electromagnetic wave propagation and absorption, as obtained from Maxwell's equations is simulated. The governing equation of the electric field wave (for a non-magnetic media) is expressed as \citep{comsol2012}
\begin{equation}
\bigtriangledown\times\mu_{r}^{-1}\left( \bigtriangledown \times E \right) -k_{0}^{2}\left( \epsilon _{r} -\frac{j \sigma}{\omega \epsilon_{0}} \right)E=0
\end{equation}
where $\omega$ is the angular frequency, $\epsilon_{0}$ is the permittivity of vacuum (8.85$\times$10$^{-12}$ F/m), $\epsilon_{r}$ is the relative permittivity, $\mu_{0}$ is the relative permeability, k$_{0}$ is the wave number in free space and $\sigma$ is the electrical conductivity. k$_{0}$ is expressed as \citep{comsol2012}
\begin{equation}
k_{0}=\omega\sqrt{\epsilon_{0}\mu_{0}}=\frac{\omega}{c_{0}}
\end{equation}
where c$_{0}$ is the speed of light in vacuum. The absorbed electromagnetic (microwave) energy is dissipated as thermal energy within the tissue phantom. The volumetric heat generation (Q) within the phantom is function of the local electric field (E) strength as \citep{datta1993}
\begin{equation}
Q= \pi f \epsilon_{0}k^{''}\vert E \vert^{2}
\end{equation}
where k$^{''}$ is the imaginary part of the dielectric constant of the phantom and \textit{f} is the frequency of the microwave radiation.

\subsection{Thermo-physical and dielectric properties of nanoparticle mixed tumor phantom }
The equivalent value for composite consisting of agar gel and injected nanoparticles depends on particles concentration and volume of nanoparticles mixed with tissue phantom. In this study for simplicity, the mean value of specific heat, density and dielectric properties for nanocomposite tissue phantom can be approximated by the respective volume properties of the two materials \citep{Wang:2010:1546-1955:1025} due to lack of experimental data. The values of thermophysical and dielectric properties are given in Table 1.
For density ($\rho$):
\begin{equation}
\rho_{mix}=\left( 1-\eta\right) \rho_{t}+\eta \rho_{np}
\end{equation}
For other properties:
\begin{equation}
C_{mix}=\left( 1-\eta\right) C_{t}+\eta C_{np}
\end{equation}
For thermal conductivity (k):
\begin{equation}
\frac{1}{k_{mix}}=\frac{(1-\eta)}{k_{t}}+\frac{\eta}{k_{np}}
\end{equation}
where,
\begin{equation}
\eta=n(\frac{1}{6} \pi d^{3}_{np})
\end{equation}
where $\eta$ is the volume concentration of nanoparticles which depend upon the number of nanoparticles per unit volume (n) and diameter of nanoparticles (d$_{np}$). The subscripts t, np, and mix refers to the tissue phantom, nanoparticles and nanoparticles composite tissue-phantom respectively. But these parametric differences between normal tissue phantom and nanocomposite tissue phantom are not apparently large because the volume concentration of particles is small. The thermophysical and dielectric properties of Al$_{2}$O$_{3}$ are taken from the literature \citep{JACOB201245}. Due to lack of robust data set for TiO$_{2}$ at 2450 MHz, the simulation is carried out for Al$_{2}$O$_{3}$ infused tumors only.  

\subsection{Boundary conditions}
An impedance boundary condition has been defined for the inner walls of the oven cavity and the waveguide. This refers that an incident wave can only penetrate into the boundary by a short distance only, and the corresponding equation is as \citep{comsol2012}
\begin{equation}
\sqrt{\frac{\mu_{0}\mu_{r}}{\epsilon_{0}\epsilon_{r}-j\frac{\sigma}{\omega}}}n\times H-\left( n.E\right) n=\left( n.E_{s}\right) n-E_{s}
\end{equation}
A perfect magnetic body boundary condition is used to define the symmetry boundaries (Eqn. (11)), expressed as
\begin{equation}
n\times H=0
\end{equation}
The microwave irradiation at 900 W is fed into the oven cavity through the port boundary condition. The port is a rectangular waveguide that operates in TE$_{10}$ mode, and excited at 2.450 GHz frequency. The port boundary condition requires a propagation constant $\beta$ which is expressed as \citep{comsol2012}
\begin{equation}
\beta=\frac{2\pi}{c_{0}}\sqrt{\nu^{2}-\nu_{c}^{2}}
\end{equation}
Where $\nu$ is the microwave frequency and $\nu_{c}$ is the cut-off frequency. Further, the cavity walls are set as adiabatic. The heat transfer analysis is considered only for the phantom. Initially, the temperature distribution within the phantom is set as uniform and is considered from the experiments as 
\begin{equation}
T\left( t_{0}\right) =25 ^{o}C
\end{equation}
The temperature continuity boundary condition \citep{comsol2012} has been imposed at the interface of the nanoparticle infused phantom and surrounding bare tissue. The outer surfaces are considered to be insulated since the experimental test section material has very low thermal conductivity.

\subsection{Numerical simulation}
All simulations in the present study have been performed using the Finite Element Method (FEM) based commercial solver COMSOL Multiphysics. The symmetry cut is applied vertically through the oven, wave-guide, phantom and turntable (refer Fig. 2(a)) to reduce the size of the model by half, such that computational time is reduced. For discretizing the computational domain, the irradiation wavelength is considered to be the governing factor for maximum the nodal space. To generate mesh size, a physics control mesh with normal element size is chosen. Simulations have been carried out using different element sizes, viz. 1/5$^{th}$, 1/10$^{th}$ and 1/15$^{th}$ of the wavelength; there negligible change in the temperature distribution has been obtained. Using a finer mesh element leads to a fractional change in temperature drop only. Thus, physics control mesh element size of 1/15$^{th}$ of the wavelength was used, resulting in 17886633 domain elements, 51633 boundary elements, and 1256 edge elements as shown in Fig. 2(b). The simulations were performed using a workstation running on two quad-core Intel Xeon, 3.50 GHz processers, along with 32 GB DDR4 RAM.\\
\begin{center}
$<$Figure 3$>$
\\
$<$Figure 4$>$
\end{center}

\section{Results and discussions}
\subsection{Transient thermal history during microwave heating}
The computationally predicted and experimentally measured temperature histories have been compared and illustrated in Fig. 3 and Fig. 4. Figure 3 represents the temporal variation of phantom temperature at R=0 and at different axial depths whereas, Fig. 4 depicts the variation at R=10 mm, and different depths. It can be observed that good agreement between the predicted and measured thermal history has been obtained in majority of the cases. Some deviation is noted for the cases of large depths, and along the central axis. The experimental deviation noticed can be attributed to the influence of possible chemical reaction during microwave heating process and heat loss from the phantom. Also, the phantom being a soft gel undergoes decrease in viscosity and structural stiffness on heating, which leads to localized changes in the thermo-physical and dielectric properties, thus leading to experimental deviation. Additionally, a linear increase in predicted temperature has been obtained with respect to time due to the assumption of no moisture loss from the phantom in the simulation. It is also noteworthy that the metallic thermocouples within the phantom and the oven cavity alter the standing waves, thus continuously changing the wave pattern of both strong and weak electric field within the oven cavity. This may also lead to the observed deviations in the thermal distribution within the system.

Figures 3 and 4 show that the maximum temperature is located near the core region (R=0 mm, 10 mm and Z=35 mm) of the phantom. The primary reason behind this is the low thermal conductivity of the tissue phantom, which allows the heat to dissipate rather slowly during 20 s of continuous microwave heating. Another possible reason could be the propagation of the microwave radiation, which is composed of nodes and antinodes of minimum and maximum heating respectively, through the tissue phantom. The position of the nodes depend upon several parameters, such as dielectric properties, specific geometry of the cavity, i.e. position of wave guide, size of sample, placement of sample into cavity \citep{Brace2007Jul, Phasukkit2009Nov, doi:10.1111/j.1365-2621.1993.tb06198.x}, etc. \\
\begin{center}
$<$Figure 5$>$
\\
$<$Figure 6$>$
\end{center}

\subsection{Spatial thermal distribution in the tissue phantom}
The temperature of tissue phantom measured at different axial positions for different radial locations is shown in Fig. 5. It is evident from the figure that for all axial positions, the maximum phantom temperature has been observed at the central axis of the phantom (R=0 mm). However, at a depth of Z=20 mm from the top surface, the maximum temperature distribution has been obtained at R=10 mm, which is away from the central axis. Again, at Z=30 mm from the phantom surface, the minimum temperature distribution has been obtained at R=10 mm.  This reflects the uneven heating patterns of tissue phantoms during MWA. The temperature variation along axial and radial positions shows wavy character, as illustrated in Fig. 6. This behavior corresponds to the absorption of microwave energy with oscillating pattern of electromagnetic wave fields. The presence of the nodes and antinodes lead to maximum or minimum temperatures, which leads to wavy nature of the thermal distribution. To prevent the excessive evaporation of moisture from the phantom (to ensure that it does not dry up during the heating), its top surface has been covered with aluminum foil. The moisture collected in the gap between foil and phantom surface is excited by the microwave irradiation and gets warmer. This result in an increase in axial temperature profile at the top half section of the phantom compared to the bottom half section. The radial temperature profile at different axial positions shows a decreasing trend towards the outer surface from the center. A relatively higher microwave heating rate at the phantom center causes an increase in temperature at the core zone. Such enhancements in core temperature of tumors will definitely be helpful for clinicians to treat subsurface tumors, which are normally difficult to reach during thermal therapy, and often lead to thermal damage of neighboring healthy tissue.

\subsection{Thermal efficacy of nanoparticle mediated microwave heating}
Having understood the thermal history and distribution during MWA of tissue phantoms, the objective now lies on increasing the core temperature to higher levels to ensure necrosis of the tumor, and to contain the heat within the tumor zone. This could be brought about by injecting nanoparticle based colloids (marked healthy by the use of bio-friendly surface moieties) into the tumor, mostly in conjunction with other drug solutions. For MWA, the nanoparticles to be employed should possess high values of the imaginary part of the dielectric constant (to convert microwave irradiation to thermal energy via polarization hysteresis). The nanoparticles should also possess high thermal conductivity (to increase thermal transport within the tumor zone and confine the same). Nanoparticles mixes with agar gel in proper proportions poured at the center of the tissue phantom and having cylindrical shape as shown in Fig. 2(c). Earlier research \citep{doi:10.1111/j.1365-2621.1993.tb06198.x, YANG2004445, doi:10.1080/08327823.1991.11688160, Pitchai2015Dec} work have showed an agreement of maximum rise in transient temperature at the central core region during continuous microwave heating. Therefore, it will be interesting and of direct utilities to understand the temperature distribution and thermal containment behavior of tissue phantom infused with dielectric nanoparticles compared to tissue without nanoparticles (W.N.P). 

The comparison of transient temperature evolution for different nanoparticle infused tissue phantom is shown in Fig. 7. TiO$_{2}$ is commonly known as photocatalyst \citep{article, B923755K} which generates free radicals in absorbance of electromagnetic waves. The free radicals make it an opaque body which act as perfect reflector in the presence of electromagnetic wave and reflect most of its energy incident upon it. Thus, result in rise in average transient temperature to the surrounding tissue phantom at Z= 30 mm and 40 mm as shown in Fig. 7(a) and (c) respectively compared to the midpoint (Z=35 mm) as shown in Fig. 7(b). The effect of nanoparticles at the central core has been found to be insignificant as compared to the tissue phantom in absence of nanoparticles. Also, the presence of TiO$_{2}$ along with tissue phantom and water molecules generates hydroxyl radicals (OH molecules) \citep{doi:10.1002/0471238961.0914151805070518.a01.pub2, kubaschewski1983titanium} by absorption of electromagnetic wave and form a membrane to the outer surface of the nanoparticles mixed tissue phantom which causes reflection of microwave energy. \\
\begin{center}
$<$Figure 7$>$
\\
$<$Figure 8$>$
\end{center} 
 
\subsection{Thermal distribution and hot spots within tissue phantom}
The predicted tissue thermal history in absence of nanoparticles during microwave heating for 20 s is shown in Fig. 8(a) and 9. It is observed that there is a formation of hotspot at the cylindrical tissue phantom. The tissue phantom has highest temperature of 80 $^{o}$C at central core (Z=35 mm) as shown in Fig. 9(d), due to higher dielectric loss and higher dielectric constant that the tissue phantom possess, which in turn leads to higher absorption of microwave energy. The higher microwave energy further efficiently converted into heat. The spatial temperature contour plots at different axis positions as demonstrated in Fig. 9 clearly shows that there is an uneven heating of phantom, which in turn results in formation of hot and cold spots. The reason behind uneven heating is mainly due to the position of the phantom in the microwave cavity \citep{Brace2007Jul} and uneven deposition of water molecules in the phantom. The existence of more water molecules in the phantom sample results in improvement in the absorption of microwave energy. A good agreement between the predicted and measured uneven heating have been already shown in Figures 3 and 4. The numerically predicted results for alumina nanoparticles infused tissue phantom is shown in Fig. 8(b) and 10. The additional heat carried by nanocomposite tissue phantom having lower specific heat and higher thermal conductivity as compared to normal tissue results in transient dissipation of the heat away from core region, such that there is immediate rise in maximum temperature up to $\backsim$139 $^{o}$C, where a good agreement between experimental and numerical results is noted. Also, as one moves away from central region the temperature distribution is very similar to normal tissue phantom. Spatial counter plots at different depths as illustrated in Fig. 10 shows the zones of hot and cold spots. This highlights the uneven heating pattern in case of microwave heating, which follows the general trend of experimental transient temperature profiles as discuss earlier. The results from the experiments and simulations indicate that unlike other modalities of nanoparticle infused hyperthermia (laser, magnetic, etc.), direct infusion of nanoparticles to the tumor leads to reduction of the tumor temperature compared to the surrounding healthy tissues. In the present case, this is caused by the dielectric loss component of the nanoparticles employed. Since the tissue phantom (and bio-tissues) is largely composed of water, the system has a large dielectric constant, with a relatively large loss component. With the addition of the nanoparticles (with alumina having low dielectric constant compared to water, and titanium oxide with marginally higher values) the scenario changes. As alumina has a relatively weaker loss component, the effective dielectric loss component within the tumor is reduced. This decreases the dielectric relaxation of the tumor phantom region, which leads to reduced heat generation within the tumor compared to the neighboring tissue.  \\
\begin{center}
$<$Figure 9$>$
\\
$<$Figure 10$>$
\end{center}

\section{Conclusion}
An experimental and computational investigation of microwave assisted thermal heating (MWATH) of tissue phantom using a domestic microwave oven has been reported in this work. Detailed experiments were performed using an in-house experimental setup, fabricated to mimic the conditions during MWA of biological tissues. Tissue phantoms synthesized from agar gels were employed in the present experiments. Computer simulations using finite element method based tools were carried out to support the experimental observations and probe insight on the thermal transport aspects deep within the tissue phantom. All simulations were carried out using the commercial solver Comsol. Good agreement between the predicted and measured temperature were achieved for the different cases studied. Furthermore, experiments were conducted to investigate the efficacy of dielectric nanoparticles viz. alumina (Al$_{2}$O$_{3}$) and titanium oxide (TiO$_{2}$) during the MWATH of nanoparticle infused tumor phantoms. A deep seated tumor injected with nanoparticle solution was specifically mimicked in the experiments. The tissue phantom was synthesized in a new approach to ensure a deep seated tumor, with nanoparticles injected into it, is mimicked, which would closely resemble a similar tumor in a patient. Interesting results were obtained in terms of spatiotemporal thermal history of the nanoparticle infused tissue phantoms. An elevation in the temperature distribution was achieved in the vicinity of the targeted zone due to the presence of nanoparticles, and the spatial distribution of temperature was grossly morphed. It has been conclusively shown, using both experiments and simulations that unlike other nanoparticle mediated hyperthermia techniques (like laser, photothermal, magnetic, etc.), direct injection of the nanoparticles within the tumor leads to enhanced heat generation in the neighboring healthy tissues.  The inhomogeneity of the hyperthermia event is evident from the local occurrence of hot spots and cold spots respectively. The present findings may have far reaching implications as a framework in predicting temperature distributions during MWA, design protocols for nano-drug mediated cancer ablation, and establishing patient-friendly clinical processes to optimize the hyperthermia treatment.



\pagebreak

\bibliographystyle{elsarticle-num-names}
\hspace{-.42cm}\textbf{References} 
\bibliography{reference}





\pagebreak
\hspace{-.6cm}\textbf{Table captions:}\\
Table 1: Properties and parameters used for simulation\\
Table 2: Geometrical dimensions of the computational domain \\
\\
\hspace{-.6cm}\textbf{Figure captions:}\\
Figure 1: ((a) Tissue phantom without nanoparticles, (b) top view of tissue phantom sample with nanoparticle infused phantom poured at the center, (c) view of test section with thermocouples, wooden fixture and turn table, (d) schematic diagram of the experimental setup.\\
Figure 2: (a) Computational domain, (b) computational domain Showing the grid mesh, (c) experimental design for the measurement of temperature in tissue phantom.\\
Figure 3: Measured and predicted temperature transients at different axial positions for R=0 mm. \\
Figure 4: Measured and predicted temperature transients at different axial positions for R=10 mm away from the center.\\
Figure 5: Variation of temperature at different radial positions; at Z= 10 mm, 20 mm, 30 mm, 35 mm, 40 mm, 50 mm and 60 mm respectively.\\
Figure 6: Maximum temperature at different (a) axial positions and (b) radial positions .\\
Figure 7: Comparisons between temperature transients of tissue phantoms with and without nanoparticles at R=0mm and 10 mm.\\
Figure 8: Simulated thermal distribution of axial mid plane of the tissue phantom after the microwave heating; (a) tissue in absence of nanoparticles, (b) tissue mixed with 1\% Al$_{2}$O$_{3}$.\\ 
Figure 9: Simulated thermal distribution at different radial planes after the microwave heating.\\
Figure 10: Simulated thermal distribution at different radial planes after the microwave heating of tissue phantom mixed with 1\% Al$_{2}$O$_{3}$.

\pagebreak
\begin{center}
\textbf{List of Tables}
\end{center}
\begin{table}[ht]
\vspace*{4cm}
\caption{}
\centering
\begin{tabular}{l l l}
\hline\noalign{\smallskip}
Properties/Parameters & \hspace{.5cm}Tissue Phantom & \hspace{.5cm}Al$_{2}$O$_{3}$ nanoparticles \\
\hline
thermal conductivity(k$_{th}$) & \hspace{.5cm}$0.636\pm 0.004 (W/m.K)^{\ddagger}$ & \hspace{.5cm}36(W/m.K)\cite{JACOB201245}\\
 density($\rho$) &  \hspace{.5cm}$1050\pm 0.016 (kg/m^{3})^{\ddagger}$ & \hspace{.5cm} 3965(kg/m$^{3}$)\cite{JACOB201245}\\
 specific heat(C$_{p}$) & \hspace{.5cm}$3929\pm 40.32 (J/kg.K)^{\ddagger}$ & \hspace{.5cm} 880(J/kg.K)\cite{JACOB201245}\\
 dielectric constant (k$^{'}$) &  \hspace{.5cm}73.6\citep{YANG2004445} & \hspace{.5cm} 10.80\cite{JACOB201245} \\
 dielectric loss (k$^{"}$) &  \hspace{.5cm}11.5\citep{YANG2004445} & \hspace{.5cm} 0.1565\cite{JACOB201245}\\
 Relative Permeability ($\mu_{r}$) &  \hspace{.5cm}1.0 & \hspace{.5cm} 1.0\\
 Nanoparticle diameter (d$_{np}$) &  \hspace{.5cm}-- & \hspace{.5cm} 15(nm)\\
 No. of Nanoparticle (n) &  \hspace{.5cm}-- & \hspace{.5cm} 5.66$\times$10$^{21}$\\
\hline
$^{\ddagger}$Measured
\end{tabular}
\end{table}

\pagebreak

\begin{table}[ht]
\vspace*{4cm}
\caption{}
\centering
\begin{tabular}{l l l l l}
\hline\noalign{\smallskip}
 & \hspace{.1cm} Length(mm) & \hspace{.1cm}Width(mm) & \hspace{.1cm}Height(mm) & \hspace{.1cm} Radius(mm)\\
 
\hline
Microwave cavity & \hspace{.1cm} 350 & \hspace{.1cm}350 & \hspace{.1cm}210 & \hspace{.1cm} --\\
 Waveguide & \hspace{.1cm} 102 & \hspace{.1cm}66 & \hspace{.1cm}24 & \hspace{.1cm} --\\
 Turntable & \hspace{.1cm} -- & \hspace{.1cm}-- & \hspace{.1cm}6 & \hspace{.1cm} 345\\
 Bio tissue & \hspace{.1cm} -- & \hspace{.1cm}-- & \hspace{.1cm}70 & \hspace{.1cm} 35\\
\hline
\end{tabular}
\end{table}

\pagebreak
\begin{center}
\textbf{List of Figures}
\end{center}
\begin{figure}[ht]
\centering
\includegraphics[scale=.5]{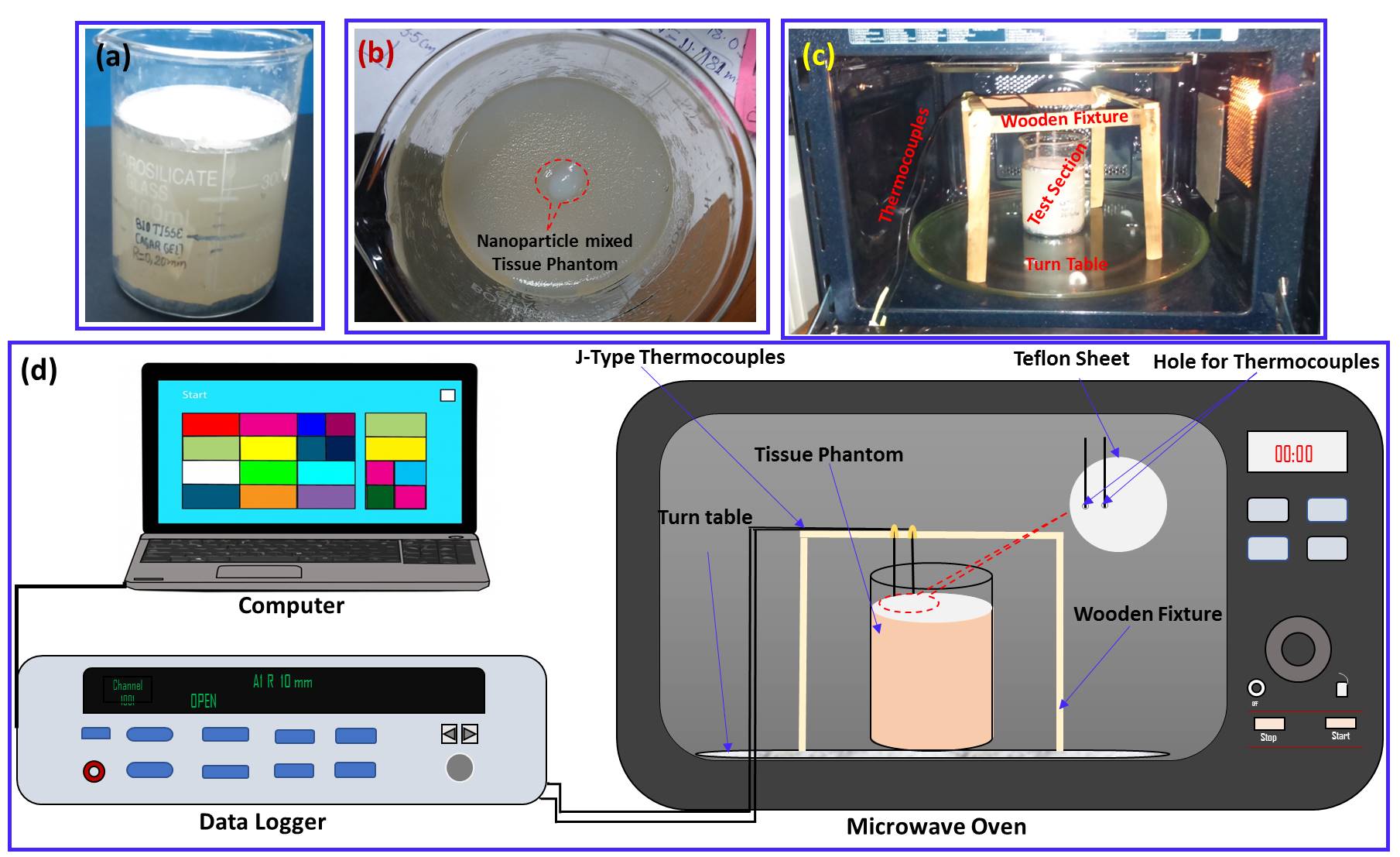}
\caption{}
\end{figure}
\pagebreak

\begin{figure}[ht]
\centering
\includegraphics[scale=.5]{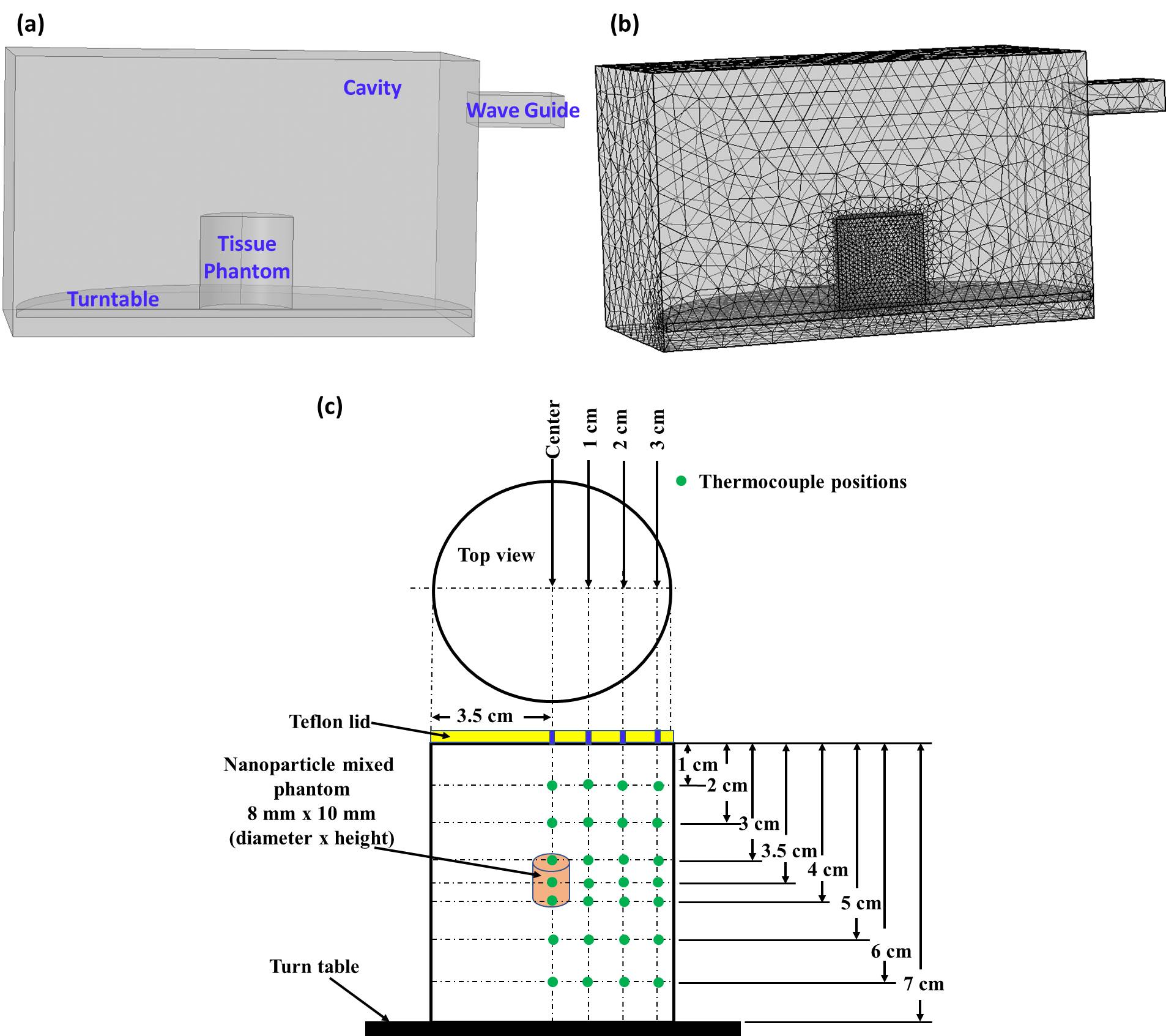}
\caption{}
\end{figure}
\pagebreak

\begin{figure}[ht]
\centering
\includegraphics[scale=.5]{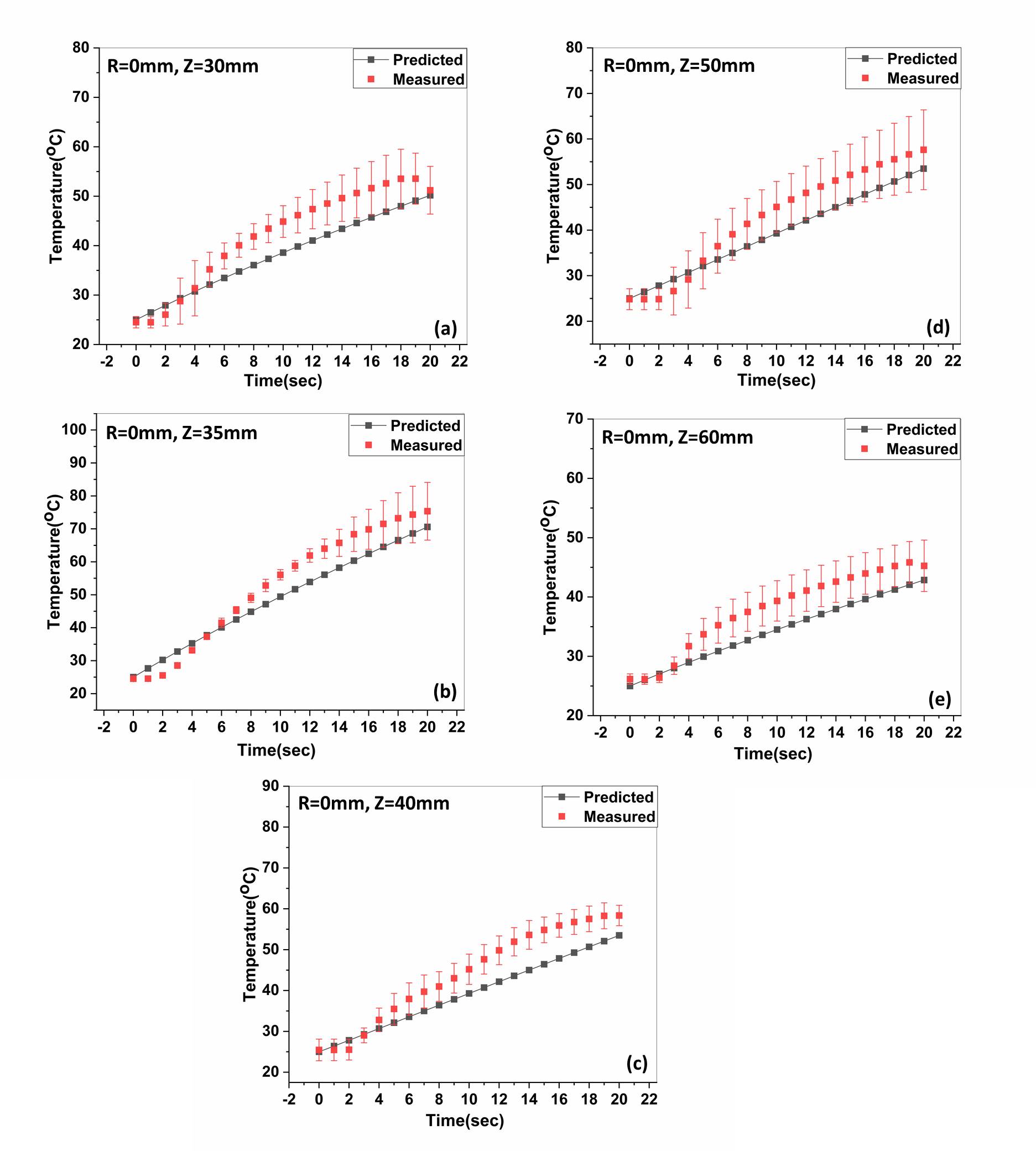}
\caption{}
\end{figure}
\pagebreak

\begin{figure}[ht]
\centering
\includegraphics[scale=.5]{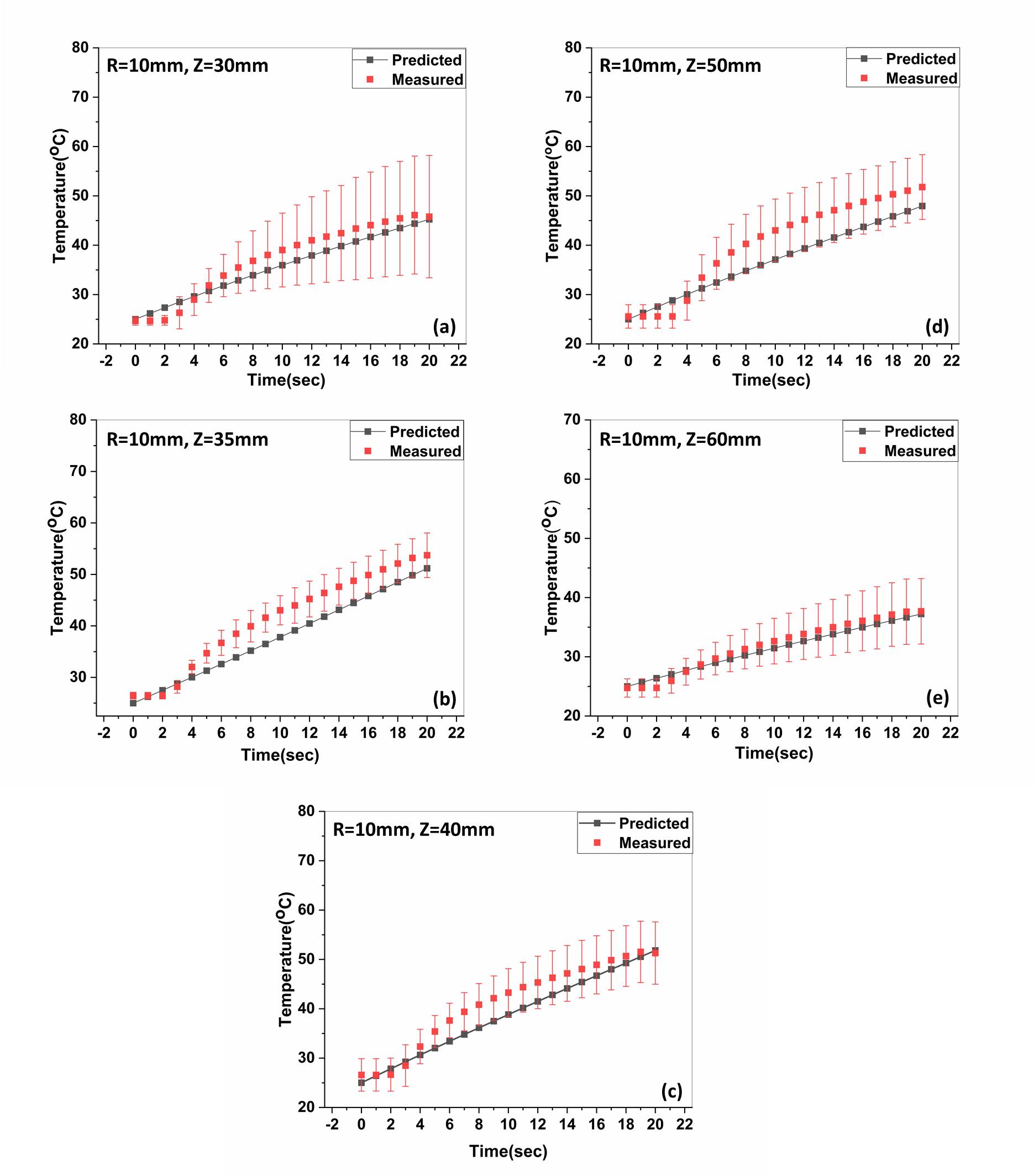}
\caption{}
\end{figure}
\pagebreak

\begin{figure}[ht]
\centering
\includegraphics[scale=.4]{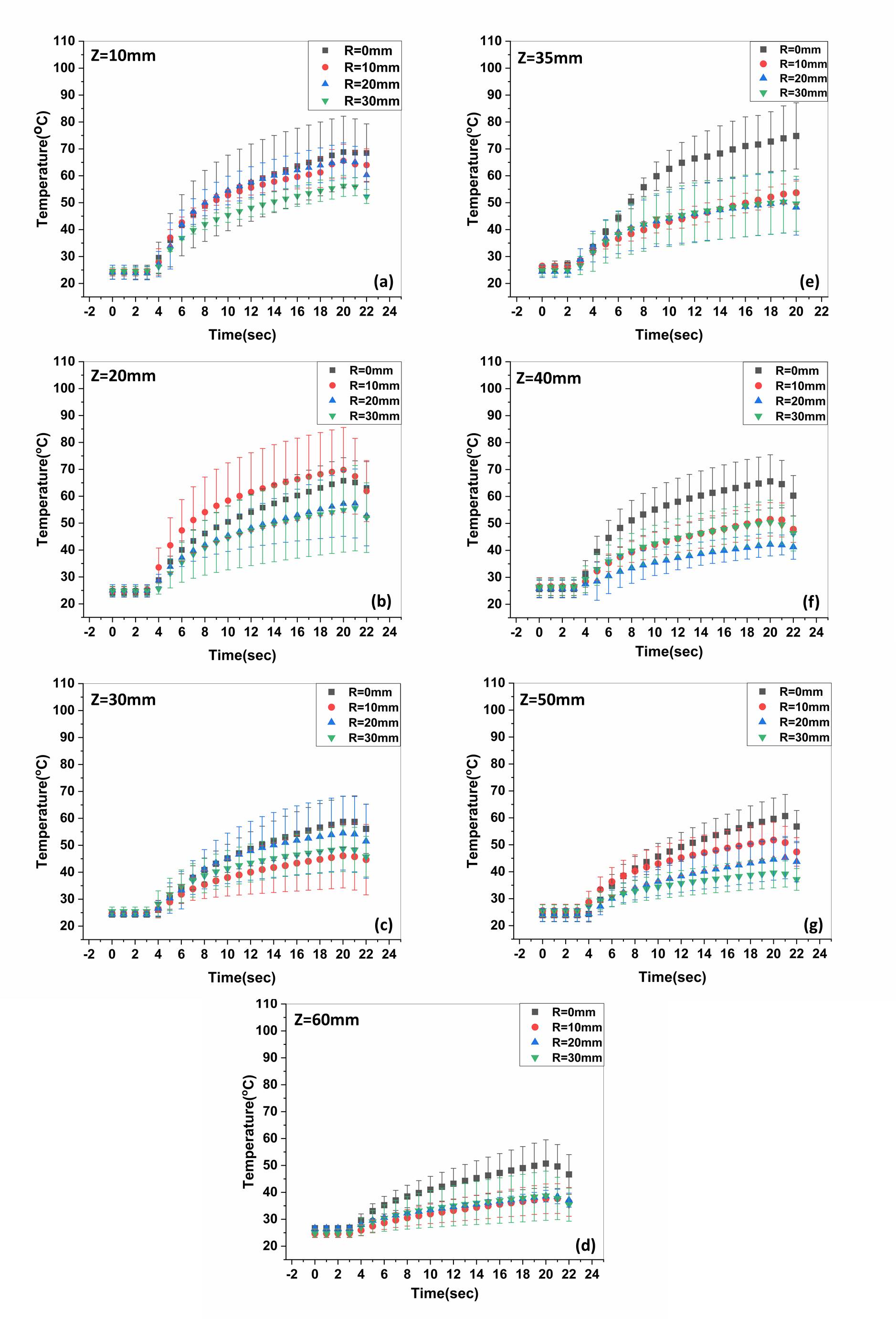}
\caption{}
\end{figure}
\pagebreak

\begin{figure}[ht]
\centering
\includegraphics[scale=.5]{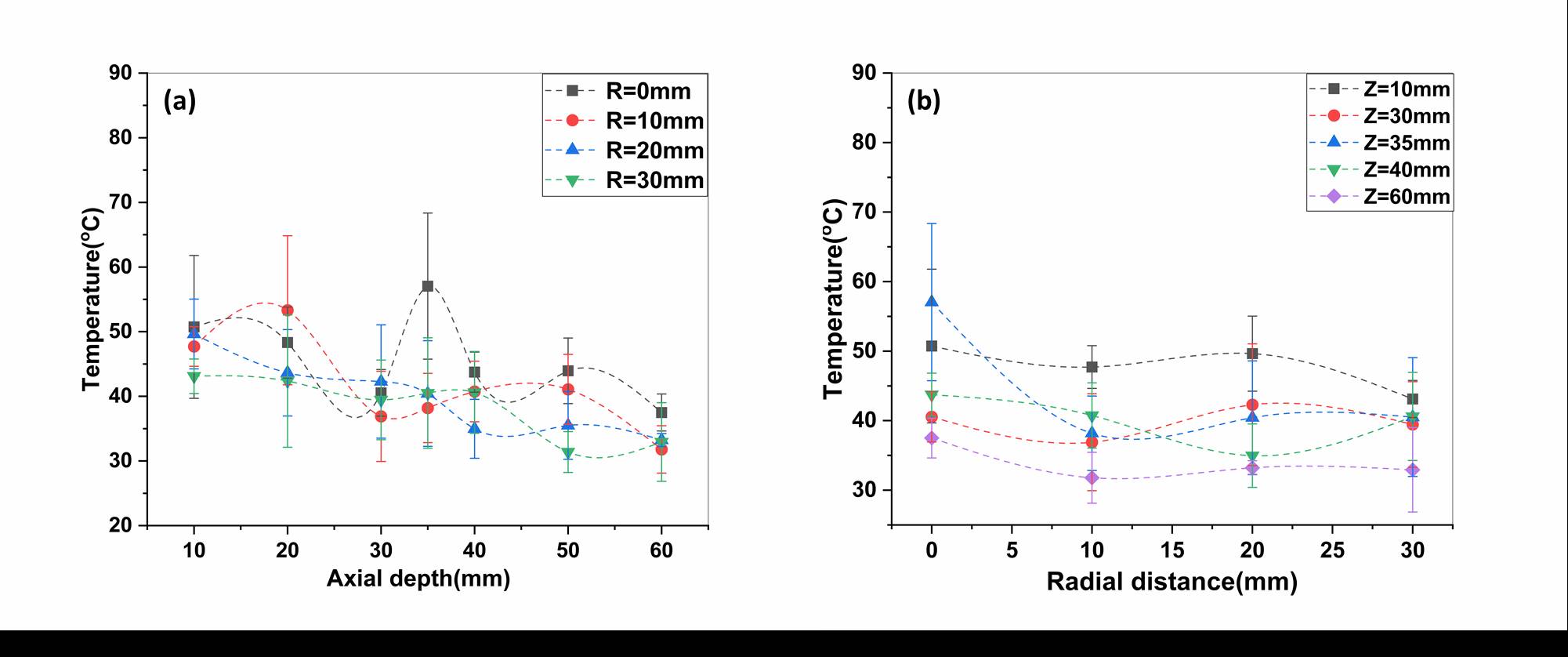}
\caption{}
\end{figure}
\pagebreak

\begin{figure}[ht]
\centering
\includegraphics[scale=.5]{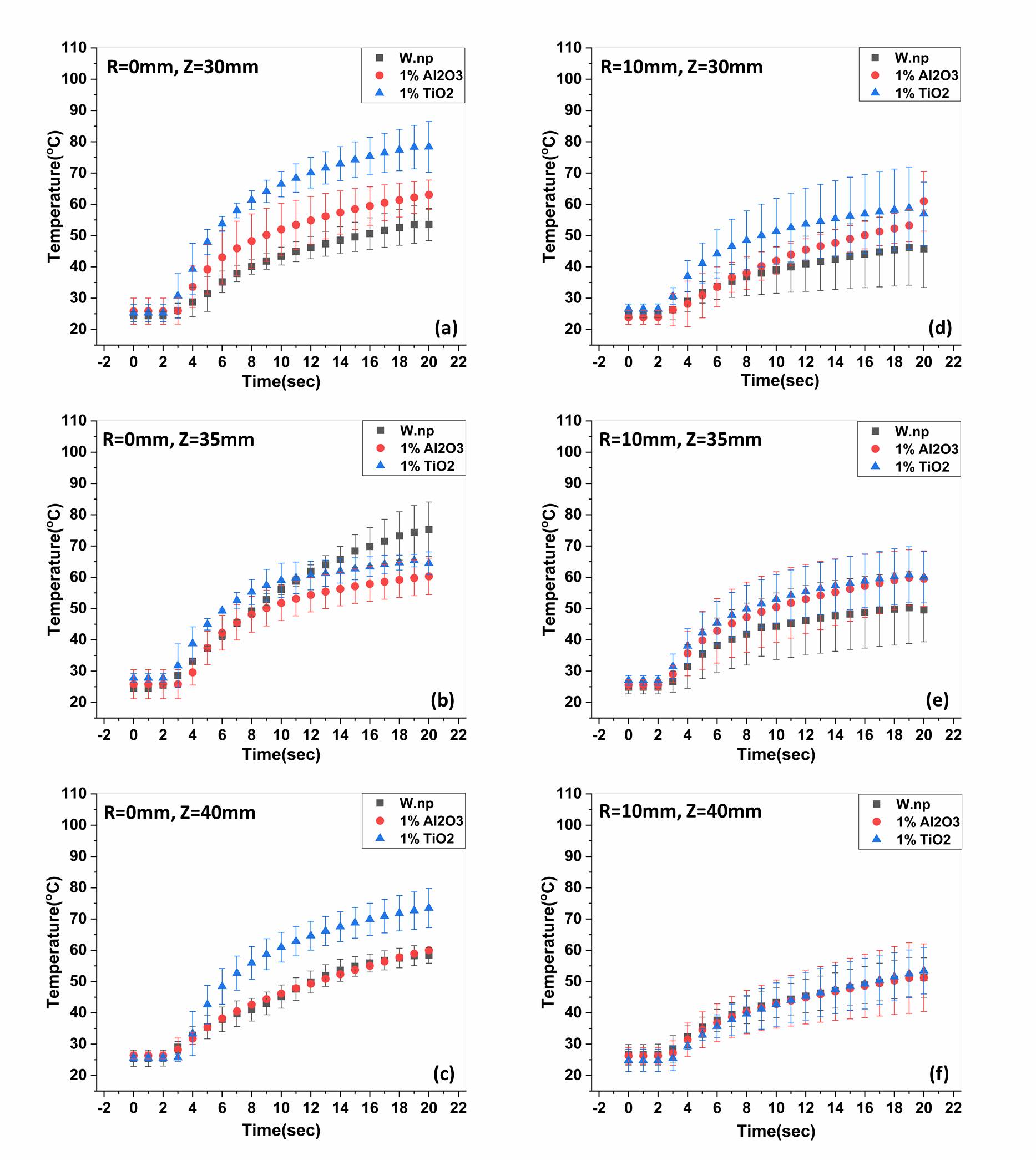}
\caption{}
\end{figure}
\pagebreak

\begin{figure}[ht]
\centering
\includegraphics[scale=.35]{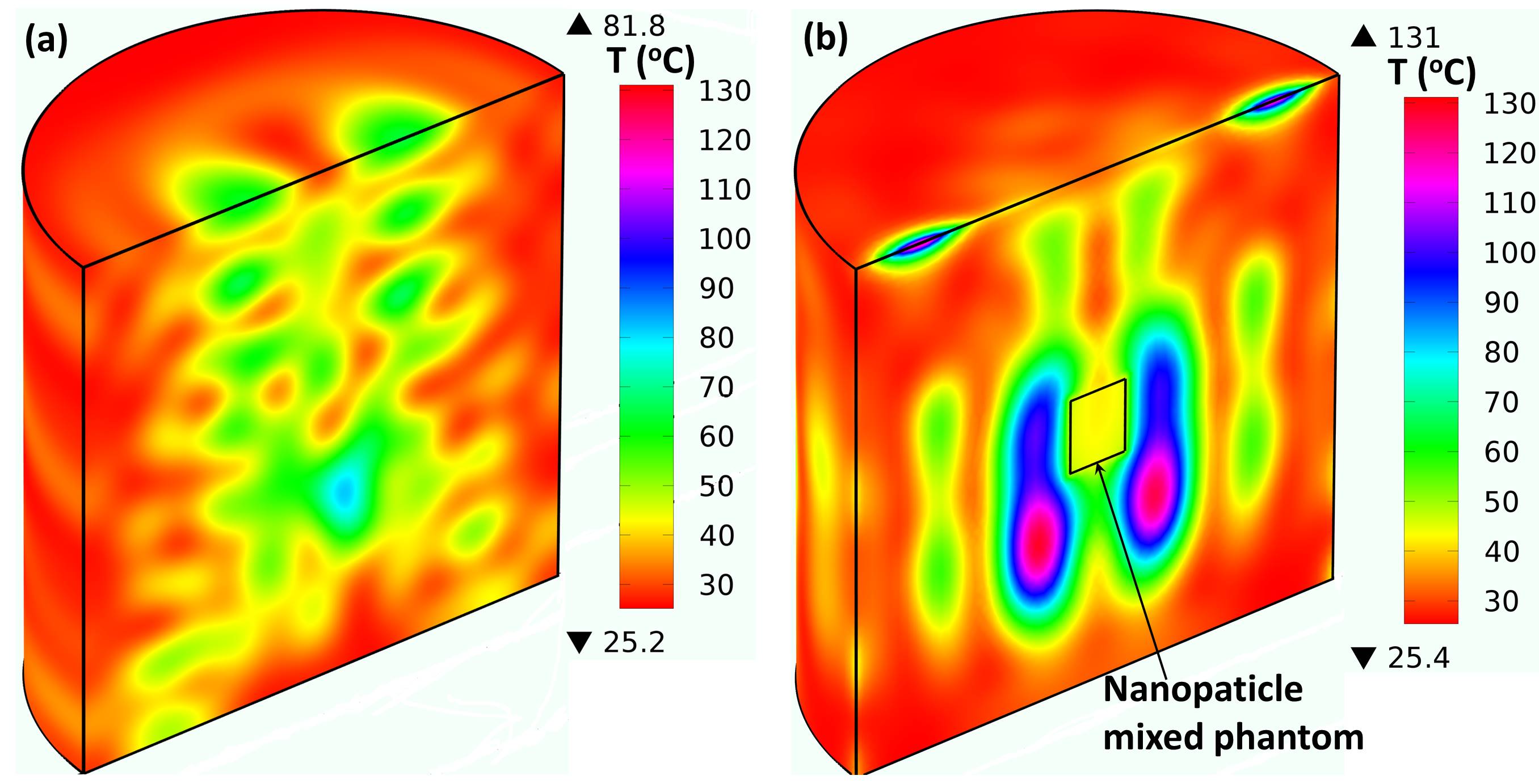}
\caption{}
\end{figure}
\pagebreak

\begin{figure}[ht]
\centering
\includegraphics[scale=.5]{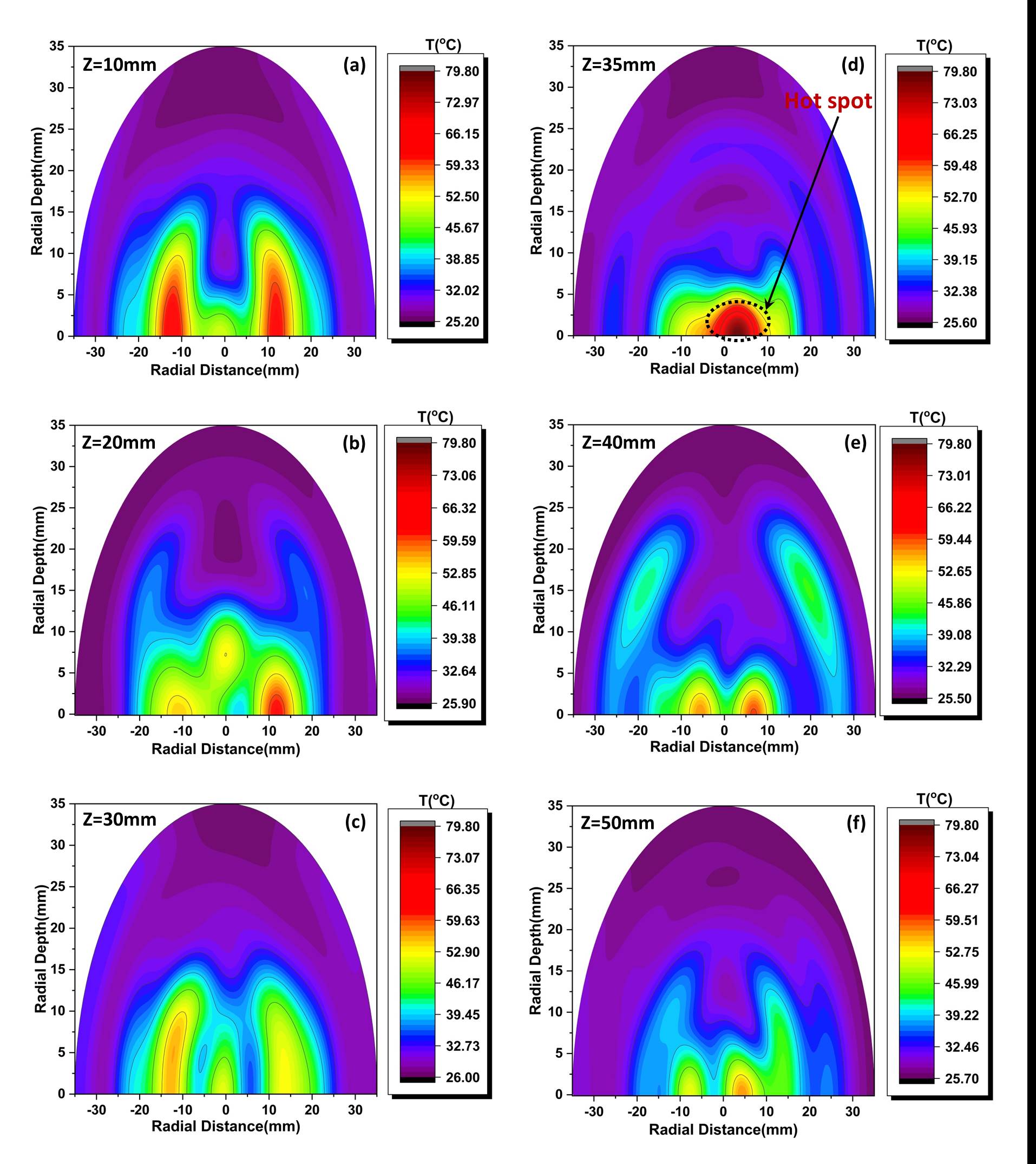}
\caption{}
\end{figure}
\pagebreak

\begin{figure}[ht]
\centering
\includegraphics[scale=.5]{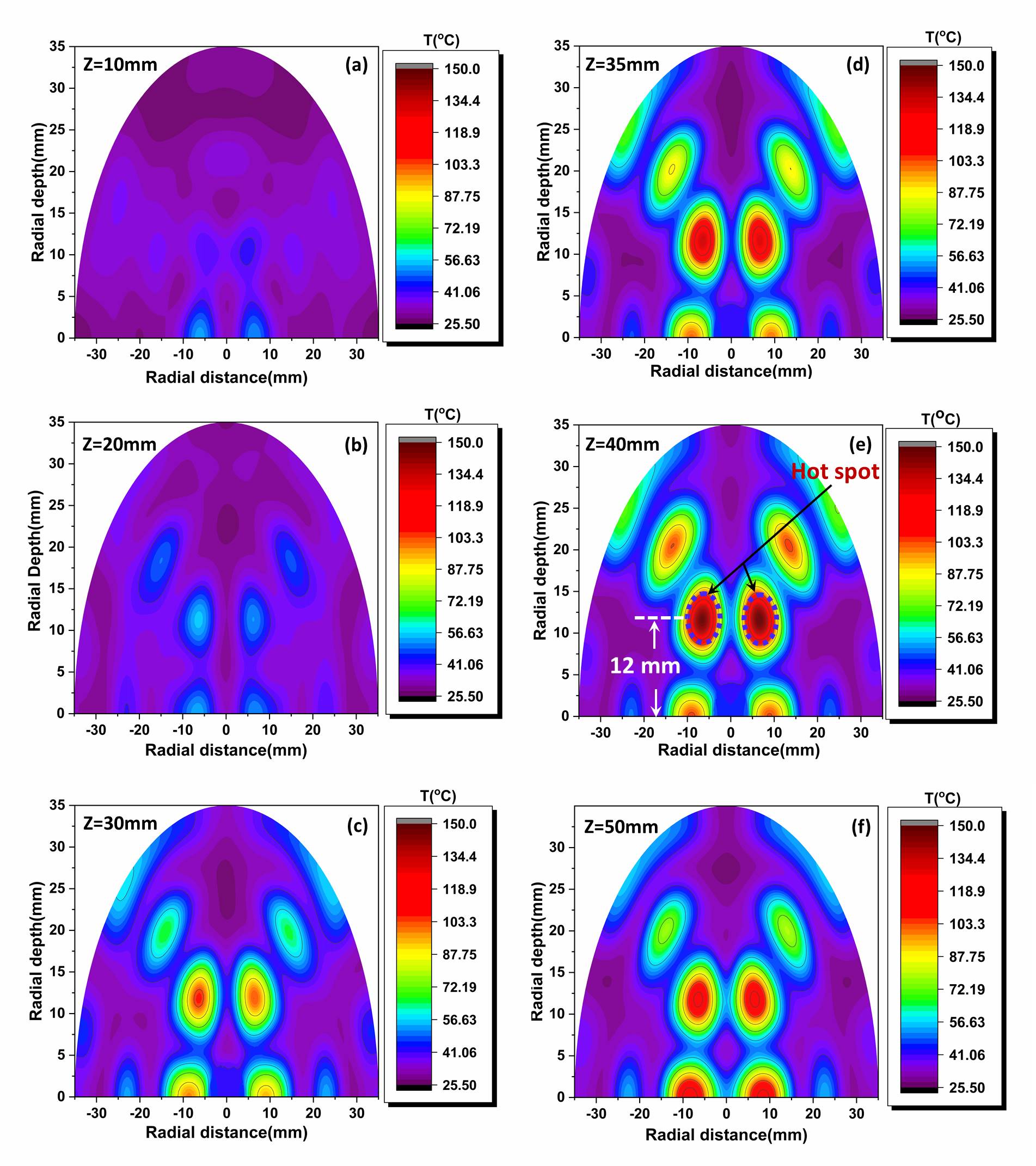}
\caption{}
\end{figure}
\pagebreak
\end{document}